\documentclass[prl,twocolumn,superscriptaddress,flushbottom,balancelastpage,a4paper,oneside,showpacs]{revtex4}

\usepackage{latexsym,amsmath,amssymb,bm,graphicx}

\makeatletter\def\Dated@name{}\makeatother

\newcommand{\mycaptionone}{%
  (Color Online) Kinks in the electronic specific heat (right) for $U/W=0.8$ (upper panel) and $1.0$ (lower panel).  The left panels show the DMFT(ED) results for the 
total energy $E_{\rm tot}$ from which the specific heat has been obtained as a numerical derivative after a spline interpolation. Also shown are
two parabolic fits (see text) valid for $T<T^*$ (red solid line) and $T>T^*$ 
(violet dotted line) and the DMFT(QMC) results of Ref.\ \onlinecite{Nils} (green dots).}

\newcommand{\mycaptiontwo}{%
  (Color Online) Kink in the low temperature specific heat of LiV$_2$O$_4$
 (open blue circles)  visible at $T^* \sim 5-6$K, and well reproduced by 
our analytical theory (red solid line)
}

\newcommand{\mycaptionthree}{%
  (Color Online) Analytical theory describing  kinks in the specific
  heat (red solid line) on the basis of the AGD formula (see text). The blue dashed line in the inset was fitted to the (red) numerical renormalization group data of Ref.\ \onlinecite{NRGT=0}; note that the deviation at larger frequencies $\omega$ does not significantly affect the specific heat in the plotted temperature range. The agreement of the analytical calculation with our numerical results (black crosses) is excellent.}
\newcommand{\myacknowledgements}{%
We acknowledge helpful discussions with R. Arita, L. Boeri,   S. B\"uhler-Paschen, K.\ Byczuk, H. Freire,  P. Hansmann, P. Jakubczyk, M.\ Kollar, J. Matsuno, W. Metzner, N. Miura, I.\ A.\ Nekrasov, G. Sangiovanni, D.\ Vollhardt, Y.-F.\ Yang and R. Zehyer.
In particular, we thank   N. Bl\"umer, R. Bulla and M. Nohara
for making available their raw data.}

\newcommand{\mybibliography}{%
  
}

\begin{document}
Ã¢â€°Ë†
  \title{Kinks in the electronic specific heat of strongly correlated systems}

  \author{A.\ Toschi}
  \affiliation{Max-Planck Institute for Solid State Research,
    Heisenbergstr.~1, 70569 Stuttgart, Germany}
   \affiliation{Institute for Solid State Physics, Vienna University of Technology,
1040 Vienna, Austria}

 \author{M. Capone} 

  \affiliation{SMC, CNR-INFM, and Dipartimento di Fisica- Universit\`a di Roma ``La Sapienza'', Piazzale Aldo Moro 2, 00185 Roma, Italy}
  \affiliation{ISC-CNR,Via dei Taurini 19, 00185 Roma Italy}

 \author{C. Castellani}
  \affiliation{SMC, CNR-INFM, and Dipartimento di Fisica- Universit\`a di Roma ``La Sapienza'', Piazzale Aldo Moro 2, 00185 Roma, Italy}

   \author{K. Held}

  \affiliation{Max-Planck Institute for Solid State Research,
    Heisenbergstr.~1, 70569 Stuttgart, Germany}
  \affiliation{Institute for Solid State Physics, Vienna University of Technology,
1040 Vienna, Austria}

\date{\today}

\begin{abstract}
We find that the heat capacity of a strongly correlated metal presents striking changes with respect to Landau Fermi liquid theory. In contrast with normal metals, where the electronic specific heat is linear at low temperature (with a $T^3$ term as a leading correction), a dynamical mean-field study of the correlated Hubbard model reveals a clear kink in the temperature dependence, marking a rapid change from a low-temperature linear behavior and a second linear regime with a reduced slope.
Experiments on LiV$2$O$_4$ support our findings, implying that correlated materials are more resistive to cooling at low $T$ than expected from the intermediate temperature behavior.  \end{abstract}

\maketitle

If we trace from low to high temperatures the specific heat capacity $c_V =\partial E/\partial T$ of a solid, it provides for a rich variety of  information. 
For a metal it increases linearly,
$c_V= \gamma_0 T$, with the prefactor $\gamma_0$ proportional to the  density of
the electronic states, i.e.,  $\gamma_0 \sim N(E_F)$. This result is also valid for correlated systems
that maintain a normal metallic behavior. In this case we can rely on
Landau's normal Fermi liquid (FL) theory\cite{Landau}, which  describes the low-energy excitations of correlated (interacting) electrons as  ``quasiparticles'' (QP) which are adiabatically connected to the  non-interacting electrons.
  As a result, only a QP renormalization factor $Z_{\rm FL}$  needs to be included in comparison to non-interacting electrons so that $c_V=  \gamma_{FL} T$ with  $\gamma_{FL}=\gamma_0 /Z_{\rm FL}$.
  This description is so universally applicable that  special attention
  is paid to any deviation occurring   in the vicinity of special
  points (e.g., Quantum Critical Points\cite{Coleman}, where the  specific heat shows a logarithmic $T$-dependence).

   Turning back  to the normal case, the common understanding \cite{Abrikosov} is that the next 
  electronic contribution to the specific heat  is cubic, $\sim T^3$. 
  This is of the same order as the contribution from the lattice degrees 
  of freedom, where the prefactor is given by the stiffness of the lattice 
  and the mass of its ions. This makes the ``lattice'' prefactor much larger than the 
  electron contribution, so that the cubic phonon contribution is
  usually dominant\cite{noteabr}. At higher temperature, finally, the specific heat saturates  with a value 
  proportional to the number of degrees of freedom in the system (law of Dulong and Petit).



  In this paper we show that the above described common understanding
  of the low-temperature specific heat of a metal needs to be markedly  corrected, if the movement of the electrons is strongly correlated
  because of their mutual Coulomb interaction. 
  Our finding is based on numerical solution of the Hubbard model using Dynamical 
  Mean-Field Theory (DMFT), combined with a field theory formula for the
  specific heat calculation given by 
  Abrikosov {\em et al.} \cite{Abrikosov} and recent results for the  energy-momentum dispersion
  relation\cite{Nekrasov06,Byczuk}.

  As mentioned above, starting point of our consideration is the half-filled single band Hubbard model,
  the minimal model which describes strongly correlated electrons on a lattice.
 This model is solved numerically  using DMFT
 \cite{Metzner,Georges} for a semicircular DOS with bandwidth $W$,  and
 exact diagonalization (ED)\cite{ed-refs} as impurity solver with $7$ energy levels in the bath.

 \begin{figure}[t!]
   \includegraphics[clip,width=80mm]{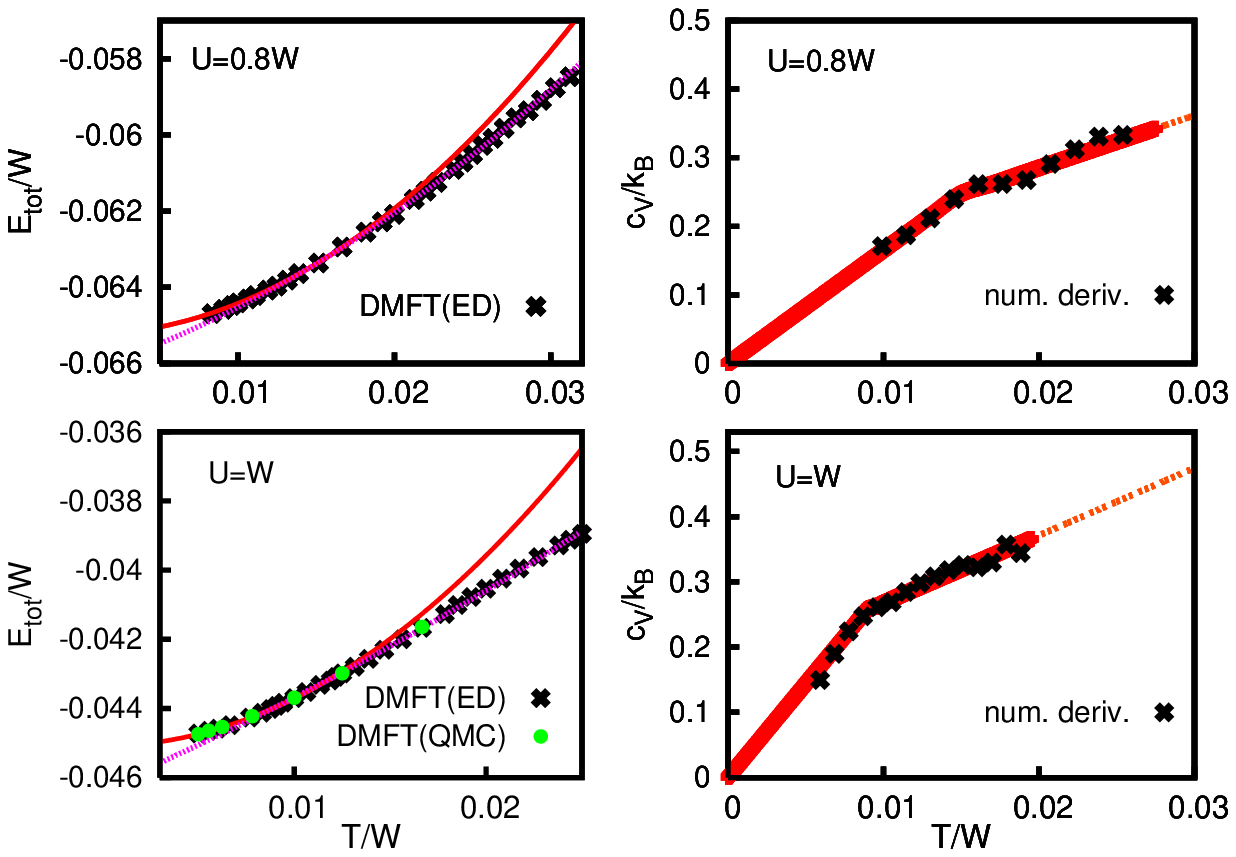}
   \caption{\mycaptionone%
   }
   \label{Fig:CvDMFT}
 \end{figure}

 Fig.\ \ref{Fig:CvDMFT} shows the total energy $E_{\rm tot}$ as a function of $T$ and the specific heat $c_V$ obtained through  numerical  differentiation for a ratio Coulomb interaction ($U$) to bandwidth ($W$) of  $U/W=0.8$ (top panels) in the temperature range where $c_V(T)$ is  monotonically increasing\cite{note}. Thanks to the extremely dense temperature mesh, our results clearly show a rapid  but continuous change of slope (kink) of $c_V$ at $T^* \sim 0.015 W$, a feature entirely unexpected for a normal FL \cite{Abrikosov}. This  kink becomes  more  and more pronounced when electronic  correlations are further enhanced by increasing the Coulomb interaction  (we show in the bottom panels of Fig.\ \ref{Fig:CvDMFT} the case  of $U/W= 1$), i.e, when moving towards the metal-to-insulator (phase) transition. At the same time, the value of $T^*$, where the kink  appears, is reduced,  displaying a clear relation with the increasing correlations which reduce $Z_{\rm FL}$.

 A proper fit to the numerical data hence needs to consist of two slopes (renormalization factors) 
$\gamma_{FL}$ and $\gamma_2$  instead of a single one:  $c_V= \gamma_{FL} T$
for $T<T^*$ and  $c_V= B +\gamma_{2} T$  for  $T>T^*$ with a rather sharp
crossover in between. This has been achieved through fitting $E_{tot}(T)=
[E_{tot}(0)+ \gamma_{FL}  \, T^2/2 \, ] \; f(T-T^*) + (E_{tot}(T^*) + BT +
\gamma_{2}T^2/2) [1-f(T-T^*)]$, using a Fermi-function-like change $f(x)=1/(1+
\mbox{e}^{\tilde{\beta}x})$ for the crossover at $T^*$. Note that this fit
 (solid red line in left panels of Fig. \ref{Fig:CvDMFT}) is
valid in the temperature range where $c_V(T)$ is monotonically
increasing, i.e., approximatively  $0 < T \lesssim 2 T^{*}$.
To assess the reliability of our impurity solver, we compared our results with precise DMFT(QMC) data \cite{Nils} (green dots in Fig. \ref{Fig:CvDMFT}, second row, first panel). 
 The comparison of the total energy shows an excellent agreement with our DMFT(ED) calculations. 
Notice that a direct observation of the kinks in DMFT(QMC) may require
a much finer grid in the temperature regime considered. In our opinion,
however, a first hint for a kink  is already provided by the  Taylor expansion  of $c_V(T)$ in Ref.\ \onlinecite{Nils}: The coefficients of the higher order terms become huge, an indication that the Taylor expansion is not appropriate as in the presence of a kink.

To support our finite-$T$ numerical findings, we carry out an analytical theory for the surprising appearance of kinks
in correlated systems, which relies on the knowledge of the $T=0$ Green
function. 
The analytical approach is based on a formula by Abrikosov, Gor'kov and Dzyaloshinski (AGD)
for the entropy of a fermionic system at low temperatures\cite{Abrikosov}. 
The AGD formula allows us to compute the entropy using the low-frequency behavior of the self energy $\Sigma(\omega)$ at zero temperature, therefore it connects the dynamical information (frequency dependence) to the thermal response (temperature dependence).
More precisely, it relates the low-temperature behavior of the entropy (and consequently of the specific heat) to the poles of the $T=0$ retarded Green function, which in turn follows from the the self-energy on the real axis.
A correlated system is expected to show a kink for a frequency $\omega^* << W$
in the $T=0$ self-energy under generic conditions \cite{Byczuk}. Also experimentally, kinks in the angular-resolved photoemission spectrum have been  observed for several strongly correlated materials such as  cuprates \cite{Lanzara}, vanadates \cite{Yoshida2005}  and ruthenates \cite{Aiura2004,Iwasawa2005}.

 For a normal metal, the AGD formula reproduces the standard Fermi-liquid result $\gamma = \gamma_0/Z_{\rm FL}$, with  $Z_{\rm FL}=[1-\frac{\partial \Sigma(\omega=0)}{ \partial \omega}]^{-1}$. 
In this paper we show that AGD formula also works beyond this linear
Fermi-liquid regime, and it actually describes a  kink in the specific
heat at a temperature $T^*$, if the proper ''kinky'' $T=0$ self energy for a correlated electron system is used.

Let us now prove this result. The specific heat can be expressed via the entropy as $c_V(T) = T \frac{d S}{d T}$. For a metallic system at low $T$, the entropy is computed according to AGD \cite{Abrikosov} as
\begin{eqnarray}
S(T) & = &  \frac{1}{2\pi i T}\int_{-\infty}^{\infty} d\epsilon N(\epsilon) \int_{-\infty}^{\infty} \, d\omega  \, \omega  \, \frac{\partial f(\omega)}{\partial \omega} \nonumber \\  & \times  & [\mbox{log}  G_{R}^{-1}(\epsilon,\omega) - \mbox{log} G_{A}^{-1}(\epsilon,\omega)],
\label{eq:Abr}
\end{eqnarray}
where $N(\epsilon)$ is the non-interacting DOS (We use a semicircular DOS $N(\epsilon)=\frac{4}{\pi D^2}\sqrt{D^2-\epsilon^2}$) with
bandwidth $W=2D$, $f(\omega)=\frac{1}{\mbox{e}^{\omega/T}+1}$ is the Fermi-Dirac distribution function
and $G_{R/A}(\epsilon)$ the retarded/advanced $T=0$ Green functions
respectively.  Eq.  (\ref{eq:Abr}) has been obtained in Ref. \cite{Abrikosov} 
by a low-temperature expansion of the Self-Energy. Introducing the auxiliary dimensionless variable 
$y=\omega/T$ ($k_B\equiv 1$), and performing a straightforward derivative 
w.r.t. $T$, the specific heat is eventually computed as  

\begin{eqnarray}
c_V(T) & = & T \frac{d S(T)}{d T} = T \frac{1}{2\pi i}\int_{-\infty}^{\infty} d\epsilon N(\epsilon) \int_{-\infty}^{\infty} \, dy  \, y  \nonumber \\  & \times  & \frac{ \mbox{e}^y}{ (\mbox{e}^y+1)^2}    [ G_{A}(\epsilon, yT) \frac{d}{dT}   G_{A}^{-1}(\epsilon, yT )  \nonumber \\  & - &  G_{R}(\epsilon, yT) \frac{d}{dT} G_{R}^{-1}(\epsilon, yT) ].
\label{eq:Abr3}
\end{eqnarray}
In the case of a FL, where just one renormalization factor 
$Z_{\rm FL}$ is present for the  low-frequency behavior of the self-energy, the standard FL formula ($c_V(T)= \gamma_0/Z_{FL} T$) is easily recovered.

\begin{figure}[t!]
    \begin{center}\includegraphics[clip,width=70mm]{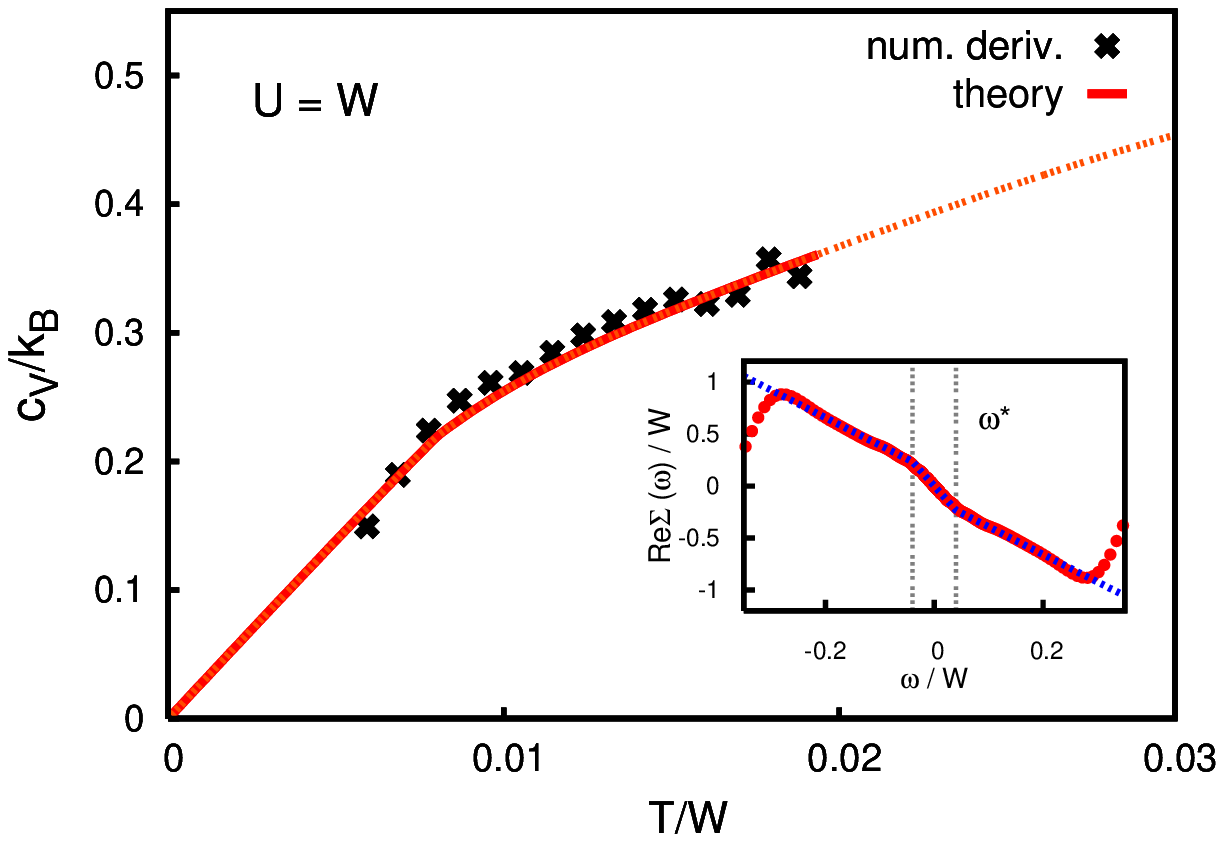}\vspace*{-5mm}\end{center}
    \caption{\mycaptionthree%
    }
    \label{Fig:AbrDMFT}
  \end{figure}

The same Eq. (\ref{eq:Abr3}), however, yields completely different results for  strongly correlated metals: When the interactions are strong enough, the spectral function displays a typical ``three-feature'' structure (the QP peak, and the two Hubbard subbands), which survives to moderate doping. In this situation
{\sl two  distinct renormalization factors} can be identified in the low-frequency regime with a kink in the real part of  $\Sigma(\omega)$ in between \cite{Byczuk}. 
Specifically, Ref. \cite{Byczuk} shows that while the lowest frequencies follow
the FL behavior Re$\Sigma(\omega)= (1-1/Z_{\rm FL}) \omega$   
 there is a rapid (but continuous) change of slope (kink) in 
Re$\Sigma(\omega)$  at frequency $\omega^*\ll W$  (e.g., $\omega^*\simeq (\sqrt{2} -1) Z_{FL} W/2$ in the case of the semicircular DOS).
For larger frequency,   Re$\Sigma(\omega)= - b + (1-1/Z_{\rm CP}) \omega$
with a reduced slope  $ Z_{\rm CP} > Z_{\rm FL}$   (typically by about a factor $2$), see inset of  Fig. \ref{Fig:AbrDMFT} and Ref.  \cite{Byczuk}. The constant $b=(1/Z_{\rm FL}- 1/Z_{\rm CP}) \omega^*$  ensures the continuity of $\Sigma(\omega)$.

As a consequence of this self energy kink, the Green functions and their temperature derivatives appearing in Eq. (\ref{eq:Abr3}) have to be written separately for the two regimes, namely $ G_{R} = (yT/Z_{\rm FL}-\epsilon+i0^{+})^{-1}$ and  $\frac{d}{dT}   G_{R}^{-1}(\epsilon, yT ) = y/Z_{\rm FL}$ for $\omega < \omega^*$, while for frequencies larger than $\omega^{*}$ one has $ G_{R} = (yT/Z_{\rm CP}-\epsilon+ b +i0^{+})^{-1}$ and  $\frac{d}{dT}   G_{R}^{-1}(\epsilon, yT ) = y/Z_{\rm CP}$ with $Z_{\rm CP} > Z_{\rm FL}$. The parameters have been extracted from fitting $\Sigma(\omega)$ of Ref. \cite{NRGT=0} (the use of the numerical renormalization group as an impurity solver allowing for very accurate low-frequency results). In the inset of Fig. \ref{Fig:AbrDMFT} we show $\Sigma(\omega)$ and the fit (blue line).
We recall in passing that it is $Z_{\rm CP}$, which controls the width of the ``quasiparticle'' peak in the interacting DOS; while $Z_{\rm FL}$ characterizes only the asymptotic properties in the limit $\omega\rightarrow 0$ (or  $T\rightarrow 0$)\cite{Byczuk}.   

The evaluation of Eq. (\ref{eq:Abr3}) has been performed by splitting explicitly the integral over $y$ in the two regions:
\begin{eqnarray}
\label{Eqcvabr}
c_V(T) &\! =\! &  T \left[\frac{1}{Z_{\rm FL}}\!\!\! \int\limits_{|y|< \frac{\omega^*}{T}}\!\!\!N(\frac{yT}{Z_{\rm FL}}) + \frac{1}{Z_{\rm CP}}\!\!\!\int\limits_{|y|> \frac{\omega^*}{T}}\!\!\! N(\frac{yT}{Z_{\rm CP}}\!+\!b) \right]  \nonumber  \\  &   &  \times  d y \,  \frac{y^2  \mbox{e}^y}{ (\mbox{e}^y+1)^2}
\end{eqnarray}
This equation is the final result of our analytical calculation. It allows us to compute, through a simple integral, the specific heat from the non-interacting density-of-states $N(E)$, the 
two renormalization factors $Z_{\rm FL}$ and $Z_{\rm CP}$ and
the kink frequency $\omega^*$. 
Using the parameters extracted from Ref. \cite{NRGT=0} we obtain the
solid line shown in Fig. \ref{Fig:AbrDMFT}. 

It is easy to verify that
the standard Fermi-liquid behavior is recovered from Eq. \ref{Eqcvabr}
in the limit of large $\omega^*$ (i.e., when only one low-frequency
scale is present). In the opposite limit $\omega^*\to 0$ a standard
Fermi-liquid behavior is also recovered, though with a different
renormalization factor $\gamma =\gamma_0/Z_{\rm CP}$. More interesting
is the intermediate situation, which we are considering here, 
when $\omega^*$ lies in
the low-frequency range. In this case, the specific heat behavior shows
a kink at a temperature $T^* \propto \omega^*$ (with a proportionality
factor of about $1/5$ for the case of the semicircular DOS): As one can
see in Fig. \ref{Fig:AbrDMFT} the standard FL behavior
$c_V(T) =\frac{\gamma_0}{Z_{\rm FL}}  T$ is recovered only for $T< T^*$,
while at $T= T^*$ a sharp change of slope is observed. For $T> T^*$, the
specific heat is still essentially linear, but with a completely different slope
(determined by the value of $Z_{\rm CP}$ and the coefficient
$b$). Eventually, the AGD formula loses its validity  at higher 
temperatures, where the maximum of $c_V$ is reached (see again \cite{note}). 

We emphasize that the AGD formula reproduces the finite low-$T$ DMFT(ED)
solution  not only qualitatively but also at a quantitative level. This allows to precisely relate the value of $T^*$ with $\omega^*$ and,  hence, with the characteristic parameters of the system (e.g., the estimate $T^* \sim 1/10 (\sqrt{2}-1) Z_{FL} W$ works well for the case of a semicircular DOS).  
The agreement with the AGD formula is particularly remarkable if we notice that, strictly speaking,  the AGD formula is only  applicable to the linear-$T$ regime, since it does not include all additional terms leading to 
the aforementioned  $T^3$ contribution.
However, this term is small at low temperatures. Hence, if the correlation is strong enough, it can push the kink in the very small $T$ regime, where the AGD formula is expected to work.
This explains why our analytical calculation is able to reproduce our numerical results to
a very good accuracy in  Fig.\ \ref{Fig:AbrDMFT}.
Let us emphasize that  it was not at all clear {\it a priori} whether an AGD-like calculation was  possible beyond the regime of Landau's QP, i.e., after the kink in the energy-momentum dispersion which indicates the basic excitations are no longer Landau QP.

The theoretical evidence of a low-temperature kink in the electronic  specific heat of strongly correlated systems poses the question of its experimental observation, which was -so far- still lacking. 
The main problem is obviously the  phonon contribution $c_V\sim T^3$ which -because of its large prefactor-  usually overshadows the much smaller electronic contribution to the specific heat,
 already at temperatures of few ten Kelvin.
 This restricts the choice to materials which show the kink at a very low  $T^*$. That 
 means in turn compounds with a strong renormalization ($Z_{\rm FL} \ll 1$), i.e., 
  heavy Fermion systems. Given our starting point, the Hubbard model, the ideal
 material is LiV$_2$O$_4$, the first d-electron system where  heavy Fermion 
 behavior was found  \cite{LiPaper}. Indeed,  recent LDA+DMFT 
calculations \cite{Arita}, which take into account the realistic three- d-band structure of LiV$_2$O$_4$, have demonstrated that an effective description in terms  of the single-band Hubbard model (very close to half-filling) is particularly appropriate for this compound.

\begin{figure}[t!]
    \begin{center}\includegraphics[clip,width=70mm]{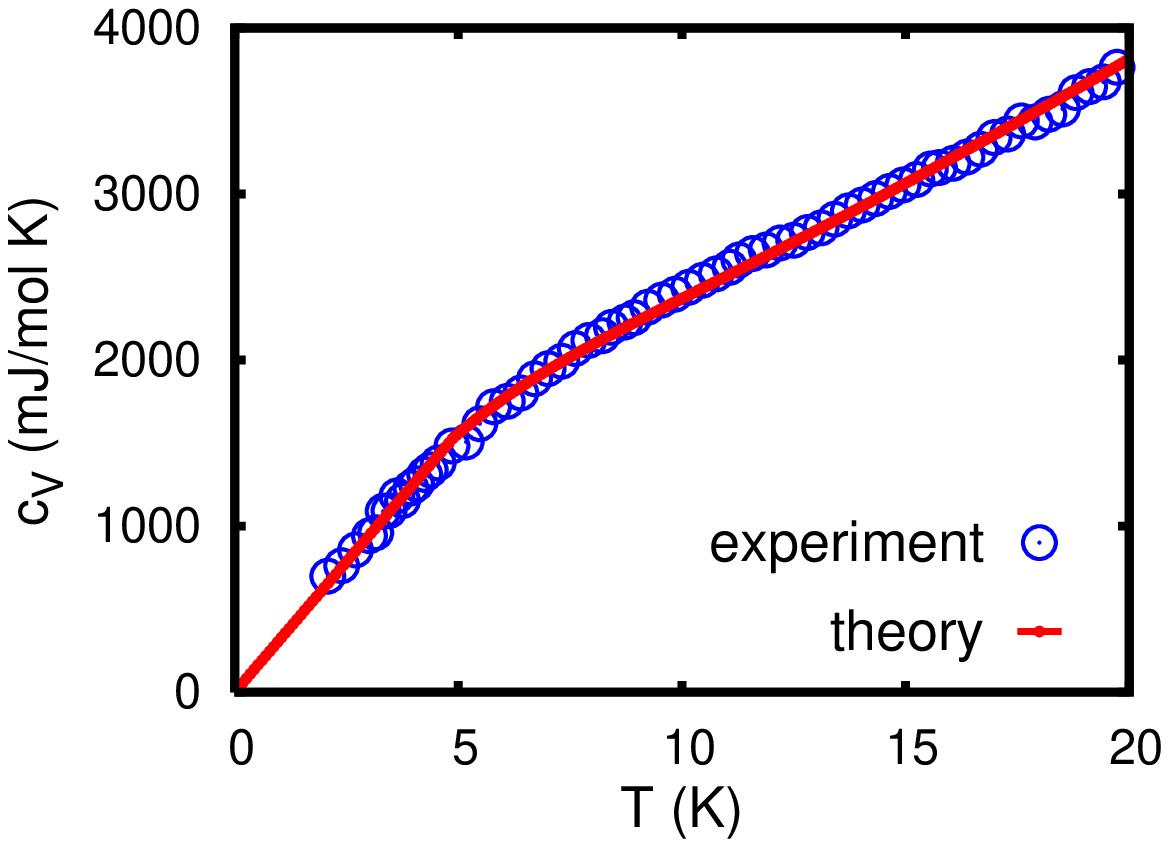}\vspace*{-5mm}\end{center}
    \caption{\mycaptiontwo%
    }
    \label{Fig:LiDMFT}
  \end{figure}

In Fig. \ \ref{Fig:LiDMFT}, we show that our theory nicely describes the experimental results for 
LiV$_2$O$_4$. We compare the data of Ref. \onlinecite{Urano} (displayed in a magnified low-$T$ range with 
respect to the original publication) with our analytical formula,
fitting the free parameters to the experimental data. Indeed, a kink is
clearly visible, as at the curve rapidly changes its slope at a
temperature $T^*$ of $5-6$K.
 The three fitting parameters ($Z_{\rm FL} = 0.054$; $Z_{\rm CP}=0.092$, $\omega^*$ $=0.0035 W $ with $W =600\,$meV for LiV$_2$O$_4$) assume very reasonable values. This clearly confirms the strong-correlation origin of the kink in the specific heat of this material.

We notice there are also kinks in the specific heat of  $f$-electron heavy Fermions such as YbRh$_2$Si$_2$\cite{Trovarelli} or YbCu$_{5-x}$Al$_x$\cite{Bauer}. However these materials are close to a quantum critical point, at which additional physical processes become important. In some 
systems also long range magnetic order leads to additional structures in the specific heat. Hence, at present, it is less clear in how far these kinks are connected to our theory.
Another material with strongly correlated Fermions showing similar kinks in the specific heat is $^3$He (Ref.\ \onlinecite{He3A,He3B}) for which  however the application of a lattice model
such as the Hubbard model represents certainly quite a crude approximation.

In conclusion, we have demonstrated numerically, analytically and experimentally that the textbook knowledge of the  electronic specific heat at low temperatures needs to be modified for strongly correlated electrons. 
In the proximity of the Mott transition the leading correction to the linear Fermi-liquid temperature behavior is a quite  rapid change of slope, i.e., a {\em kink}, which takes place well before (at smaller $T$)  the  standard $T^3$  behavior becomes relevant.
  Let us emphasize the reported kink is a generic feature of strongly correlated 
electron systems, in very contrast to existing theories for 
  kinks stemming from the coupling to (potentially present) bosonic degrees of freedom .
Since the slope of the specific heat is  reduced after the kink, the behavior of $c_V/T$ is just opposite to what one would expect from the standard theory, i.e., $c_V/T$ is  decreasing
with increasing temperature instead of the expected increase due to the cubic term.
Hence, if one extrapolates from the behavior at intermediate temperatures (i.e., after the kink) without taking into account the kink, a much lower  specific heat at low temperatures is obtained with respect to the actual result. In other words, a material with strongly correlated electrons can be unexpectedly  resistant against cooling at low temperatures.
Moreover, depending on the temperature range considered in the experiments, only one of the two regimes of linear behavior of $c_V(T)$ may be accessible. This can easily lead to remarkable inconsistencies in the analysis of the experimental data for strongly correlated materials.

\myacknowledgements

\mybibliography

\onecolumngrid


\begin{thebibliography}{10}


\bibitem{Landau}
     L. Landau,
JEPT {\bf 3}, 920 (1957).


\bibitem{Coleman} 
P.~Coleman, and A.~J.~Schofield,
{ Nature} {\bf 433}, 226 (2005).


\bibitem{Abrikosov} 
  A.~A.~Abrikosov, L.~P.~Gor'kov, and I.~E.~.Dzyaloshinski, Methods of Quantum Field Theory in Statistical Physics. { Dover Publications, New York} (1963).


\bibitem{noteabr} 
 Strictly speaking
 the leading correction to the linear behavior
 is a (relatively small) contribution  $\sim T^3 \log(1/T)$, stemming from 
 the coupling between electronic and lattice (or other bosonic) degrees of 
 freedom \cite{Abrikosov}.
 



  \bibitem{Nekrasov06} I.~A.~Nekrasov {\em et al.},
{ Phys.\ Rev.\ B} {\bf 73}, 155112 (2006).

\bibitem{Byczuk} 
 K.~Byczuk {\em et al.},
{ Nature Physics} {\bf 3}, 168 (2007). 


  \bibitem{Metzner}  W.~Metzner and D.~Vollhardt,
{ Phys.\ Rev.\ Lett.} {\bf 62}, 324 (1989).

  \bibitem{Georges} A.~Georges  {\sl et al.},
    { Rev.\ Mod.\ Phys.} {\bf 68}, 13 (1996).


\bibitem{ed-refs} M.~Caffarel and W.~Krauth, Phys. Rev. Lett. {\bf{72}}, 1545 (1994).

\bibitem{note}
        For these values of $U$, $c_V$ displays a maximum\cite{Georges} which 
        marks the upper limit 
        of the temperature region where coherent metallic processes 
        are predominant. In Fig. \ref{Fig:CvDMFT} (right panels) we show data points below this maximum. 

\bibitem{Nils} 
N.~Bl\"umer,
{ Phys.\ Rev.\ B} {\bf 76},  205120 (2007). 


\bibitem{Lanzara}
    A.~Lanzara {\em et al.},
{ Nature} {\bf 412}, 510 (2001).

  \bibitem{Yoshida2005}
    T.~Yoshida {\em et al.},
{ Phys.\ Rev.\ Lett.} {\bf 95}, 146404 (2005); {\it ibid.} {\bf 44}, 187 (1995).

  \bibitem{Aiura2004}
     Y.~Aiura {\em et al.},
    { Phys.\ Rev.\ Lett.} {\bf 93}, 117005 (2004).

  \bibitem{Iwasawa2005}
   H.~Iwasawa {\em et al.}, 
{ Phys. Rev. B} {\bf 72}, 104514 (2005).

\bibitem {NRGT=0} R.~Bulla,  
{ Phys.\ Rev.\ Lett.\ } \textbf{83}, 136 (1999).

\bibitem{LiPaper}
S.~Kondo  {\sl et al.}, 
{ Phys.\ Rev.\ Lett.\ } {\bf 78}, 3729 (1997).

\bibitem{Arita} 
R.~Arita {\sl et al.},
{ Phys.\ Rev.\ Lett.\ } {\bf 98}, 166402 (2007).


\bibitem{Urano}
C.~Urano {\em et al.},
{ Phys.\ Rev.\ Lett.\ } {\bf 85}, 1052 (2000).

\bibitem{Trovarelli} 
O.~Trovarelli {\em et al.},
{ Phys. Rev. Lett.} {\bf 85}, 626 (2000).  




\bibitem{Bauer} E.~Bauer {\sl et al.}, Phys. Rev. B {\bf 60}, 1238 (1999). 


\bibitem{He3A}
D.~Greywall,
{ Phys. Rev. B} {\bf 27}, 2747 (1983). 

\bibitem{He3B} Seiler {\sl et al.}
{ J. Low Temp. Phys.} {\bf 64}, 195 (1986).

  \end{thebibliography}
\end{document}